%
%
%
\def\unredoffs{} \def\redoffs{\voffset=-.31truein\hoffset=-.48truein}
\def\speclscape{}
%
%
%
%
%
\newbox\leftpage \newdimen\fullhsize \newdimen\hstitle \newdimen\hsbody
\tolerance=1000\hfuzz=2pt
\catcode`\@=11 
\ifx\hyperdef\UNd@FiNeD\def\hyperdef#1#2#3#4{#4}\def\hyperref#1#2#3#4{#4}\fi
\def\bigans{b }
\def\answ{b }
%
\ifx\answ\bigans\message{(This will come out unreduced.}
\magnification=1200\unredoffs\baselineskip=16pt plus 2pt minus 1pt
\hsbody=\hsize \hstitle=\hsize 
\else\message{(This will be reduced.} \let\l@r=L
\magnification=1000\baselineskip=16pt plus 2pt minus 1pt \vsize=7truein
\redoffs \hstitle=8truein\hsbody=4.75truein\fullhsize=10truein\hsize=\hsbody
\output={\ifnum\pageno=0 
  \shipout\vbox{\speclscape{\hsize\fullhsize\makeheadline}
    \hbox to \fullhsize{\hfill\pagebody\hfill}}\advancepageno
  \else
  \almostshipout{\leftline{\vbox{\pagebody\makefootline}}}\advancepageno
  \fi}
\def\almostshipout#1{\if L\l@r \count1=1 \message{[\the\count0.\the\count1]}
      \global\setbox\leftpage=#1 \global\let\l@r=R
 \else \count1=2
  \shipout\vbox{\speclscape{\hsize\fullhsize\makeheadline}
      \hbox to\fullhsize{\box\leftpage\hfil#1}}  \global\let\l@r=L\fi}
\fi
%
\newcount\yearltd\yearltd=\year\advance\yearltd by -2000

\def\Title#1#2{\nopagenumbers\abstractfont\hsize=\hstitle\rightline{#1}%
\vskip 1in\centerline{\titlefont #2}\abstractfont\vskip .5in\pageno=0}
\def\Date#1{\vfill\leftline{#1}\tenpoint\supereject\global\hsize=\hsbody%
\footline={\hss\tenrm\hyperdef\hypernoname{page}\folio\folio\hss}}%
%

\def\draftmode{\message{ DRAFTMODE }\def\draftdate{{\rm preliminary draft:
\number\month/\number\day/\number\yearltd\ \ \hourmin}}%
\headline={\hfil\draftdate}\writelabels\baselineskip=20pt plus 2pt minus 2pt
 {\count255=\time\divide\count255 by 60 \xdef\hourmin{\number\count255}
  \multiply\count255 by-60\advance\count255 by\time
  \xdef\hourmin{\hourmin:\ifnum\count255<10 0\fi\the\count255}}}
\def\nolabels{\def\wrlabeL##1{}\def\eqlabeL##1{}\def\reflabeL##1{}}
\def\writelabels{\def\wrlabeL##1{\leavevmode\vadjust{\rlap{\smash%
{\line{{\escapechar=` \hfill\rlap{\sevenrm\hskip.03in\string##1}}}}}}}%
\def\eqlabeL##1{{\escapechar-1\rlap{\sevenrm\hskip.05in\string##1}}}%
\def\reflabeL##1{\noexpand\llap{\noexpand\sevenrm\string\string\string##1}}}
\nolabels
%
\global\newcount\secno \global\secno=0
\global\newcount\meqno \global\meqno=1
\def\s@csym{}
\def\newsec#1{\global\advance\secno by1%
{\toks0{#1}\message{(\the\secno. \the\toks0)}}%
\global\subsecno=0\eqnres@t\let\s@csym\secsym\xdef\secn@m{\the\secno}\noindent
{\bf\hyperdef\hypernoname{section}{\the\secno}{\the\secno.} #1}%
\writetoca{{\string\hyperref{}{section}{\the\secno}{\it\the\secno.}} {{\it #1} }}%
\par\nobreak\medskip\nobreak}
\def\eqnres@t{\xdef\secsym{\the\secno.}\global\meqno=1\bigbreak\bigskip}
\def\sequentialequations{\def\eqnres@t{\bigbreak}}\xdef\secsym{}
\global\newcount\subsecno \global\subsecno=0
\def\subsec#1{\global\advance\subsecno by1%
{\toks0{#1}\message{(\s@csym\the\subsecno. \the\toks0)}}%
\ifnum\lastpenalty>9000\else\bigbreak\fi       \global\subsubsecno=0
\noindent{\it\hyperdef\hypernoname{subsection}{\secn@m.\the\subsecno}%
{\secn@m.\the\subsecno.} #1}\writetoca{\string\quad
{\string\hyperref{}{subsection}{\secn@m.\the\subsecno}{\secn@m.\the\subsecno.}}
{#1}}\par\nobreak\medskip\nobreak}
\def\appendix#1#2{\global\meqno=1\global\subsecno=0\xdef\secsym{\hbox{#1.}}%
\bigbreak\bigskip\noindent{\bf Appendix \hyperdef\hypernoname{appendix}{#1}%
{#1.} #2}{\toks0{(#1. #2)}\message{\the\toks0}}%
\xdef\s@csym{#1.}\xdef\secn@m{#1}%
\writetoca{\string\hyperref{}{appendix}{#1}{{\it Appendix} {\it #1.}} {\it #2}}%
\par\nobreak\medskip\nobreak}
%
%
\def\checkm@de#1#2{\ifmmode{\def\f@rst##1{##1}\hyperdef\hypernoname{equation}%
{#1}{#2}}\else\hyperref{}{equation}{#1}{#2}\fi}
\def\eqnn#1{\DefWarn#1\xdef #1{(\noexpand\relax\noexpand\checkm@de%
{\s@csym\the\meqno}{\secsym\the\meqno})}%
\wrlabeL#1\writedef{#1\leftbracket#1}\global\advance\meqno by1}
\def\f@rst#1{\c@t#1a\em@ark}\def\c@t#1#2\em@ark{#1}
\def\eqna#1{\DefWarn#1\wrlabeL{#1$\{\}$}%
\xdef #1##1{(\noexpand\relax\noexpand\checkm@de%
{\s@csym\the\meqno\noexpand\f@rst{##1}}{\hbox{$\secsym\the\meqno##1$}})}
\writedef{#1\numbersign1\leftbracket#1{\numbersign1}}\global\advance\meqno by1}
\def\eqn#1#2{\DefWarn#1%
\xdef #1{(\noexpand\hyperref{}{equation}{\s@csym\the\meqno}%
{\secsym\the\meqno})}$$#2\eqno(\hyperdef\hypernoname{equation}%
{\s@csym\the\meqno}{\secsym\the\meqno})\eqlabeL#1$$%
\writedef{#1\leftbracket#1}\global\advance\meqno by1}
\def\xeqn{\expandafter\xe@n}\def\xe@n(#1){#1}
\def\xeqna#1{\expandafter\xe@n#1}
\def\eqns#1{(\e@ns #1{\hbox{}})}
\def\e@ns#1{\ifx\UNd@FiNeD#1\message{eqnlabel \string#1 is undefined.}%
\xdef#1{(?.?)}\fi{\let\hyperref=\relax\xdef\next{#1}}%
\ifx\next\em@rk\def\next{}\else%
\ifx\next#1\xeqn#1\else\def\n@xt{#1}\ifx\n@xt\next#1\else\xeqna#1\fi
\fi\let\next=\e@ns\fi\next}

\def\DefWarn#1{\ifx\UNd@FiNeD#1\else
\immediate\write16{*** WARNING: the label \string#1 is already defined ***}\fi}
%
\newskip\footskip\footskip14pt plus 1pt minus 1pt 
\def\footnotefont{\ninepoint}\def\f@t#1{\footnotefont #1\@foot}
\def\f@@t{\baselineskip\footskip\bgroup\footnotefont\aftergroup\@foot\let\next}
\setbox\strutbox=\hbox{\vrule height9.5pt depth4.5pt width0pt}
\global\newcount\ftno \global\ftno=0
\def\foot{\global\advance\ftno by1\def\foot@rg{\hyperref{}{footnote}%
{\the\ftno}{\the\ftno}\xdef\foot@rg{\noexpand\hyperdef\noexpand\hypernoname%
{footnote}{\the\ftno}{\the\ftno}}}\footnote{$^{\foot@rg}$}}
%
\newwrite\ftfile
\def\footend{\def\foot{\global\advance\ftno by1\chardef\wfile=\ftfile
\hyperref{}{footnote}{\the\ftno}{$^{\the\ftno}$}%
\ifnum\ftno=1\immediate\openout\ftfile=\jobname.fts\fi%
\immediate\write\ftfile{\noexpand\smallskip%
\noexpand\item{\noexpand\hyperdef\noexpand\hypernoname{footnote}
{\the\ftno}{f\the\ftno}:\ }\pctsign}\findarg}%
\def\footatend{\vfill\eject\immediate\closeout\ftfile{\parindent=20pt
\centerline{\bf Footnotes}\nobreak\bigskip\input \jobname.fts }}}
\def\footatend{}
%
%
\global\newcount\refno \global\refno=1
\newwrite\rfile
\def\ref{[\hyperref{}{reference}{\the\refno}{\the\refno}]\nref}
\def\nref#1{\DefWarn#1%
\xdef#1{[\noexpand\hyperref{}{reference}{\the\refno}{\the\refno}]}%
\writedef{#1\leftbracket#1}%
\ifnum\refno=1\immediate\openout\rfile=\jobname.refs\fi
\chardef\wfile=\rfile\immediate\write\rfile{\noexpand\item{[\noexpand\hyperdef%
\noexpand\hypernoname{reference}{\the\refno}{\the\refno}]\ }%
\reflabeL{#1\hskip.31in}\pctsign}\global\advance\refno by1\findarg}
\def\findarg#1#{\begingroup\obeylines\newlinechar=`\^^M\pass@rg}
{\obeylines\gdef\pass@rg#1{\writ@line\relax #1^^M\hbox{}^^M}%
\gdef\writ@line#1^^M{\expandafter\toks0\expandafter{\striprel@x #1}%
\edef\next{\the\toks0}\ifx\next\em@rk\let\next=\endgroup\else\ifx\next\empty%
\else\immediate\write\wfile{\the\toks0}\fi\let\next=\writ@line\fi\next\relax}}
\def\striprel@x#1{} \def\em@rk{\hbox{}}
\def\lref{\begingroup\obeylines\lr@f}
\def\lr@f#1#2{\DefWarn#1\gdef#1{\let#1=\UNd@FiNeD\ref#1{#2}}\endgroup\unskip}

\def\addref#1{\immediate\write\rfile{\noexpand\item{}#1}} 
\def\listrefs{\footatend\vfill\supereject\immediate\closeout\rfile\writestoppt
\baselineskip=\footskip\centerline{{\bf References}}\bigskip{\parindent=20pt%
\frenchspacing\escapechar=` \input \jobname.refs\vfill\eject}\nonfrenchspacing}
\def\startrefs#1{\immediate\openout\rfile=\jobname.refs\refno=#1}
\def\xref{\expandafter\xr@f}\def\xr@f[#1]{#1}
\def\refs#1{\count255=1[\r@fs #1{\hbox{}}]}
\def\r@fs#1{\ifx\UNd@FiNeD#1\message{reflabel \string#1 is undefined.}%
\nref#1{need to supply reference \string#1.}\fi%
\vphantom{\hphantom{#1}}{\let\hyperref=\relax\xdef\next{#1}}%
\ifx\next\em@rk\def\next{}%
\else\ifx\next#1\ifodd\count255\relax\xref#1\count255=0\fi%
\else#1\count255=1\fi\let\next=\r@fs\fi\next}
%

%
\newwrite\ffile\global\newcount\figno \global\figno=1
\def\fig{fig.~\hyperref{}{figure}{\the\figno}{\the\figno}\nfig}
\def\nfig#1{\DefWarn#1%
\xdef#1{fig.~\noexpand\hyperref{}{figure}{\the\figno}{\the\figno}}%
\writedef{#1\leftbracket fig.\noexpand~\xfig#1}%
\ifnum\figno=1\immediate\openout\ffile=\jobname.figs\fi\chardef\wfile=\ffile%
{\let\hyperref=\relax
\immediate\write\ffile{\noexpand\medskip\noexpand\item{Fig.\ %
\noexpand\hyperdef\noexpand\hypernoname{figure}{\the\figno}{\the\figno}. }
\reflabeL{#1\hskip.55in}\pctsign}}\global\advance\figno by1\findarg}
\def\listfigs{\vfill\eject\immediate\closeout\ffile{\parindent40pt
\baselineskip14pt\centerline{{\bf Figure Captions}}\nobreak\medskip
\escapechar=` \input \jobname.figs\vfill\eject}}
\def\xfig{\expandafter\xf@g}\def\xf@g fig.\penalty\@M\ {}
\def\figs#1{figs.~\f@gs #1{\hbox{}}}
\def\f@gs#1{{\let\hyperref=\relax\xdef\next{#1}}\ifx\next\em@rk\def\next{}\else
\ifx\next#1\xfig #1\else#1\fi\let\next=\f@gs\fi\next}
\def\figin{\epsfcheck\figin}\def\figins{\epsfcheck\figins}
\def\epsfcheck{\ifx\epsfbox\UNd@FiNeD
\message{(NO epsf.tex, FIGURES WILL BE IGNORED)}
\gdef\figin##1{\vskip2in}\gdef\figins##1{\hskip.5in}
\else\message{(FIGURES WILL BE INCLUDED)}%
\gdef\figin##1{##1}\gdef\figins##1{##1}\fi}
\def\DefWarn#1{}
\def\figinsert{\goodbreak\midinsert}
\def\ifig#1#2#3{\DefWarn#1\xdef#1{Fig.~\noexpand\hyperref{}{figure}%
{\the\figno}{\the\figno}}\writedef{#1\leftbracket fig.\noexpand~\xfig#1}%
\figinsert\figin{\centerline{#3}}\medskip\centerline{\vbox{\baselineskip12pt
\advance\hsize by -1truein\noindent\wrlabeL{#1=#1}\footnotefont%
{\bf Fig.~\hyperdef\hypernoname{figure}{\the\figno}{\the\figno}:} #2}}
\bigskip\endinsert\global\advance\figno by1}
\newwrite\lfile
{\escapechar-1\xdef\pctsign{\string\%}\xdef\leftbracket{\string\{}
\xdef\rightbracket{\string\}}\xdef\numbersign{\string\#}}
\def\writedefs{\immediate\openout\lfile=\jobname.defs \def\writedef##1{%
{\let\hyperref=\relax\let\hyperdef=\relax\let\hypernoname=\relax
 \immediate\write\lfile{\string\def\string##1\rightbracket}}}}%
\def\writestop{\def\writestoppt{\immediate\write\lfile{\string\pageno
 \the\pageno\string\startrefs\leftbracket\the\refno\rightbracket
 \string\def\string\secsym\leftbracket\secsym\rightbracket
 \string\secno\the\secno\string\meqno\the\meqno}\immediate\closeout\lfile}}
\def\writestoppt{}\def\writedef#1{}
\def\seclab#1{\DefWarn#1%
\xdef #1{\noexpand\hyperref{}{section}{\the\secno}{\the\secno}}%
\writedef{#1\leftbracket#1}\wrlabeL{#1=#1}}
\def\subseclab#1{\DefWarn#1%
\xdef #1{\noexpand\hyperref{}{subsection}{\secn@m.\the\subsecno}%
{\secn@m.\the\subsecno}}\writedef{#1\leftbracket#1}\wrlabeL{#1=#1}}
\def\applab#1{\DefWarn#1%
\xdef #1{\noexpand\hyperref{}{appendix}{\secn@m}{\secn@m}}%
\writedef{#1\leftbracket#1}\wrlabeL{#1=#1}}
\newwrite\tfile \def\writetoca#1{}
\def\leaderfill{\leaders\hbox to 1em{\hss.\hss}\hfill}
\def\writetoc{\immediate\openout\tfile=\jobname.toc
   \def\writetoca##1{{\edef\next{\write\tfile{\noindent ##1
   \string\leaderfill {\string\hyperref{}{page}{\noexpand\number\pageno}%
                       {\noexpand\number\pageno}} \par}}\next}}}
\newread\ch@ckfile
\def\listtoc{\immediate\closeout\tfile\immediate\openin\ch@ckfile=\jobname.toc
\ifeof\ch@ckfile\message{no file \jobname.toc, no table of contents this pass}%
\else\closein\ch@ckfile\centerline{\bf Contents}\nobreak\medskip%
{\baselineskip=18.5pt  \footnotefont
\parskip=2pt\catcode`\@=12\input\jobname.toc
\catcode`\@=12\bigbreak\bigskip}\fi}
\catcode`\@=12 
%
\edef\tfontsize{\ifx\answ\bigans scaled\magstep3\else scaled\magstep4\fi}
\font\titlerm=cmr10 \tfontsize \font\titlerms=cmr7 \tfontsize
\font\titlermss=cmr5 \tfontsize \font\titlei=cmmi10 \tfontsize
\font\titleis=cmmi7 \tfontsize \font\titleiss=cmmi5 \tfontsize
\font\titlesy=cmsy10 \tfontsize \font\titlesys=cmsy7 \tfontsize
\font\titlesyss=cmsy5 \tfontsize \font\titleit=cmti10 \tfontsize
\skewchar\titlei='177 \skewchar\titleis='177 \skewchar\titleiss='177
\skewchar\titlesy='60 \skewchar\titlesys='60 \skewchar\titlesyss='60
\def\titlefont{\def\rm{\fam0\titlerm}
\textfont0=\titlerm \scriptfont0=\titlerms \scriptscriptfont0=\titlermss
\textfont1=\titlei \scriptfont1=\titleis \scriptscriptfont1=\titleiss
\textfont2=\titlesy \scriptfont2=\titlesys \scriptscriptfont2=\titlesyss
\textfont\itfam=\titleit \def\it{\fam\itfam\titleit}\rm}
 \ifx\answ\bigans\else scaled\magstep1\fi
\ifx\answ\bigans\def\abstractfont{\tenpoint}\else
\font\absit=cmti10 scaled \magstep1
\font\abssl=cmsl10 scaled \magstep1
\font\absrm=cmr10 scaled\magstep1 \font\absrms=cmr7 scaled\magstep1
\font\absrmss=cmr5 scaled\magstep1 \font\absi=cmmi10 scaled\magstep1
\font\absis=cmmi7 scaled\magstep1 \font\absiss=cmmi5 scaled\magstep1
\font\abssy=cmsy10 scaled\magstep1 \font\abssys=cmsy7 scaled\magstep1
\font\abssyss=cmsy5 scaled\magstep1 \font\absbf=cmbx10 scaled\magstep1
\skewchar\absi='177 \skewchar\absis='177 \skewchar\absiss='177
\skewchar\abssy='60 \skewchar\abssys='60 \skewchar\abssyss='60
\def\abstractfont{\def\rm{\fam0\absrm}
\textfont0=\absrm \scriptfont0=\absrms \scriptscriptfont0=\absrmss
\textfont1=\absi \scriptfont1=\absis \scriptscriptfont1=\absiss
\textfont2=\abssy \scriptfont2=\abssys \scriptscriptfont2=\abssyss
\textfont\itfam=\absit \def\it{\fam\itfam\absit}\def\footnotefont{\tenpoint}%
\textfont\slfam=\abssl \def\sl{\fam\slfam\abssl}%
\textfont\bffam=\absbf \def\bf{\fam\bffam\absbf}\rm}\fi
\def\tenpoint{\def\rm{\fam0\tenrm}
\textfont0=\tenrm \scriptfont0=\sevenrm \scriptscriptfont0=\fiverm
\textfont1=\teni  \scriptfont1=\seveni  \scriptscriptfont1=\fivei
\textfont2=\tensy \scriptfont2=\sevensy \scriptscriptfont2=\fivesy
\textfont\itfam=\tenit \def\it{\fam\itfam\tenit}\def\footnotefont{\ninepoint}%
\textfont\bffam=\tenbf \def\bf{\fam\bffam\tenbf}\def\sl{\fam\slfam\tensl}\rm}
\font\ninerm=cmr9 \font\sixrm=cmr6 \font\ninei=cmmi9 \font\sixi=cmmi6
\font\ninesy=cmsy9 \font\sixsy=cmsy6 \font\ninebf=cmbx9
\font\nineit=cmti9 \font\ninesl=cmsl9 \skewchar\ninei='177
\skewchar\sixi='177 \skewchar\ninesy='60 \skewchar\sixsy='60
\def\ninepoint{\def\rm{\fam0\ninerm}
\textfont0=\ninerm \scriptfont0=\sixrm \scriptscriptfont0=\fiverm
\textfont1=\ninei \scriptfont1=\sixi \scriptscriptfont1=\fivei
\textfont2=\ninesy \scriptfont2=\sixsy \scriptscriptfont2=\fivesy
\textfont\itfam=\ninei \def\it{\fam\itfam\nineit}\def\sl{\fam\slfam\ninesl}%
\textfont\bffam=\ninebf \def\bf{\fam\bffam\ninebf}\rm}
%
%
\def\noblackbox{\overfullrule=0pt}
\hyphenation{anom-aly anom-alies coun-ter-term coun-ter-terms}
\def\inv{^{\raise.15ex\hbox{${\scriptscriptstyle -}$}\kern-.05em 1}}

\def\Dsl{\,\raise.15ex\hbox{/}\mkern-13.5mu D} 
\def\dsl{\raise.15ex\hbox{/}\kern-.57em\partial}

\def\lspace{\ifx\answ\bigans{}\else\qquad\fi}
\def\lbspace{\ifx\answ\bigans{}\else\hskip-.2in\fi} 
\def\boxeqn#1{\vcenter{\vbox{\hrule\hbox{\vrule\kern3pt\vbox{\kern3pt
	\hbox{${\displaystyle #1}$}\kern3pt}\kern3pt\vrule}\hrule}}}
\def\mbox#1#2{\vcenter{\hrule \hbox{\vrule height#2in
		\kern#1in \vrule} \hrule}}  
%

\def\darr#1{\raise1.5ex\hbox{$\leftrightarrow$}\mkern-16.5mu #1}

\def\roughly#1{\raise.3ex\hbox{$#1$\kern-.75em\lower1ex\hbox{$\sim$}}}

\global\newcount\subsubsecno \global\subsubsecno=0
\def\subsubsec#1{\global\advance\subsubsecno by1%
{\toks0{#1}\message{(\the\secno\the\subsecno\the\subsubsecno. \the\toks0)}}%
\ifnum\lastpenalty>9000\else\bigbreak\fi
\noindent{\it\hyperdef\hypernoname{subsubsection}{\the\secno.\the\subsecno\the\subsubsecno}%
{\the\secno.\the\subsecno.\the\subsubsecno.} #1}
\par\nobreak\medskip\nobreak}
\def\boxit#1{\vbox{\hrule\hbox{\vrule\kern8pt
\vbox{\hbox{\kern8pt}\hbox{\vbox{#1}}\hbox{\kern8pt}}
\kern8pt\vrule}\hrule}}
\def\mathboxit#1{\vbox{\hrule\hbox{\vrule\kern8pt\vbox{\kern8pt
\hbox{$\displaystyle #1$}\kern8pt}\kern8pt\vrule}\hrule}}
\def\slashchar#1{\setbox0=\hbox{$#1$}           
   \dimen0=\wd0                                 
   \setbox1=\hbox{/} \dimen1=\wd1               
   \ifdim\dimen0>\dimen1                        
      \rlap{\hbox to \dimen0{\hfil/\hfil}}      
      #1                                        
   \else                                        
      \rlap{\hbox to \dimen1{\hfil$#1$\hfil}}   
      /                                         
   \fi}
\def\sqr#1#2{{\vcenter{\vbox{\hrule height.#2pt
         \hbox{\vrule width.#2pt height#1pt \kern#1pt
            \vrule width.#2pt}
         \hrule height.#2pt}}}}


\input amssym.def
\input amssym.tex
\noblackbox
\baselineskip=14.5pt

\def\comment#1{{}}

\newif\ifnref

\nreffalse

\input epsf

\def\figin{\epsfcheck\figin}\def\figins{\epsfcheck\figins}
\def\epsfcheck{\ifx\epsfbox\UnDeFiNeD
\message{(NO epsf.tex, FIGURES WILL BE IGNORED)}
\gdef\figin##1{\vskip2in}\gdef\figins##1{\hskip.5in}
\else\message{(FIGURES WILL BE INCLUDED)}%
\gdef\figin##1{##1}\gdef\figins##1{##1}\fi}
\def\DefWarn#1{}
\def\figinsert{\goodbreak\midinsert}  
\def\ifig#1#2#3{\DefWarn#1\xdef#1{Fig.~\the\figno}
\writedef{#1\leftbracket fig.\noexpand~\the\figno}%
\figinsert\figin{\centerline{#3}}\medskip\centerline{\vbox{\baselineskip12pt
\advance\hsize by -1truein\noindent\footnotefont\centerline{{\bf
Fig.~\the\figno}\ \sl #2}}}
\bigskip\endinsert\global\advance\figno by1}

\def\iifig#1#2#3#4{\DefWarn#1\xdef#1{Fig.~\the\figno}
\writedef{#1\leftbracket fig.\noexpand~\the\figno}%
\figinsert\figin{\centerline{#4}}\medskip\centerline{\vbox{\baselineskip12pt
\advance\hsize by -1truein\noindent\footnotefont\centerline{{\bf
Fig.~\the\figno}\ \ \sl #2}}}\smallskip\centerline{\vbox{\baselineskip12pt
\advance\hsize by -1truein\noindent\footnotefont\centerline{\ \ \ \sl #3}}}
\bigskip\endinsert\global\advance\figno by1}


\def\tilde{\widetilde}


\def\br{\hfill\break}

\def\ra {\rightarrow}





\lref\StiebergerHBA{
  S.~Stieberger and T.R.~Taylor,
 ``Closed String Amplitudes as Single-Valued Open String Amplitudes,''
Nucl.\ Phys.\ B {\bf 881}, 269 (2014).
[arXiv:1401.1218 [hep-th]].
}

\lref\BernQJ{
  Z.~Bern, J.J.M.~Carrasco and H.~Johansson,
``New Relations for Gauge-Theory Amplitudes,''
Phys.\ Rev.\ D {\bf 78}, 085011 (2008).
[arXiv:0805.3993 [hep-ph]].
}

\lref\StiebergerHQ{
  S.~Stieberger,
``Open \& Closed vs. Pure Open String Disk Amplitudes,''
[arXiv:0907.2211 [hep-th]].
}

\lref\stnew{
  S.~Stieberger and T.R.~Taylor, in preparation.}

\lref\KawaiXQ{
  H.~Kawai, D.C.~Lewellen and S.H.H.~Tye,
``A Relation Between Tree Amplitudes Of Closed And Open Strings,''
  Nucl.\ Phys.\  B {\bf 269}, 1 (1986).
}
\lref\CachazoXEA{
  F.~Cachazo, S.~He and E.~Y.~Yuan,
[arXiv:1412.3479 [hep-th]].}
\lref\CachazoNSA{
  F.~Cachazo, S.~He and E.~Y.~Yuan,
  ``Einstein-Yang-Mills Scattering Amplitudes From Scattering Equations,''
[arXiv:1409.8256 [hep-th]].}

\lref\CachazoIEA{
  F.~Cachazo, S.~He and E.~Y.~Yuan,
  ``Scattering of Massless Particles: Scalars, Gluons and Gravitons,''
JHEP {\bf 1407}, 033 (2014).
[arXiv:1309.0885 [hep-th]].}

\lref\CachazoHCA{
  F.~Cachazo, S.~He and E.~Y.~Yuan,
  ``Scattering of Massless Particles in Arbitrary Dimensions,''
Phys.\ Rev.\ Lett.\  {\bf 113}, 171601 (2014)
[arXiv:1307.2199 [hep-th]].}

\lref\CachazoGNA{
  F.~Cachazo, S.~He and E.~Y.~Yuan,
Phys.\ Rev.\ D {\bf 90}, no. 6, 065001 (2014).
[arXiv:1306.6575 [hep-th]].}

\lref\Bohr{
  N.E.J.~Bjerrum-Bohr, P.H.~Damgaard, T.~Sondergaard and P.~Vanhove,
 ``The Momentum Kernel of Gauge and Gravity Theories,''
JHEP {\bf 1101}, 001 (2011).
[arXiv:1010.3933 [hep-th]].
}

\lref\BernSV{
  Z.~Bern, L.J.~Dixon, M.~Perelstein and J.S.~Rozowsky,
``Multileg one loop gravity amplitudes from gauge theory,''
Nucl.\ Phys.\ B {\bf 546}, 423 (1999).
[hep-th/9811140].
}

\lref\SiegelSK{
  W.~Siegel,
  ``Hidden gravity in open string field theory,''
Phys.\ Rev.\ D {\bf 49}, 4144 (1994).
[hep-th/9312117].
}

\lref\StiebergerCEA{
  S.~Stieberger and T.~R.~Taylor,
  ``Graviton as a Pair of Collinear Gauge Bosons,''
Phys.\ Lett.\ B {\bf 739}, 457 (2014).
[arXiv:1409.4771 [hep-th]].}

\lref\BrittoFQ{
  R.~Britto, F.~Cachazo, B.~Feng and E.~Witten,
  ``Direct proof of tree-level recursion relation in Yang-Mills theory,''
Phys.\ Rev.\ Lett.\  {\bf 94}, 181602 (2005).
[hep-th/0501052].
}

\lref\notation{M.L.~Mangano and S.J.~Parke,
``Multiparton amplitudes in gauge theories,''
Phys. Rept.  {\bf 200}, 301 (1991).
[hep-th/0509223];\br
L.J.~Dixon,
  ``Calculating scattering amplitudes efficiently,''
in Boulder 1995, QCD and beyond 539-582.
[hep-ph/9601359].}

\lref\sakh{A.D.\ Sakharov, ``Vacuum Quantum Fluctuations in Curved Space
and the Theory of Gravitation,''   Dokl.\ Akad.\ Nauk SSSR {\bf 177}, 70 (1967) [Gen.\ Rel.\ Grav.\ {\bf 32}, 365 (2000)].}

\lref\Veneziano{D.~Amati and G.~Veneziano, ``Metric From Matter,''
Phys.\ Lett.\ B {\bf 105}, 358 (1981);\br
S.L.~Adler,
  ``Einstein Gravity as a Symmetry Breaking Effect in Quantum Field Theory,''
Rev.\ Mod.\ Phys.\  {\bf 54}, 729 (1982), [Erratum-ibid.\  {\bf 55}, 837 (1983)].}

\lref\KosteleckyPX{
  V.A.~Kostelecky, O.~Lechtenfeld and S.~Samuel,
``Covariant String Amplitudes On Exotic Topologies To One Loop,''
Nucl.\ Phys.\ B {\bf 298}, 133 (1988).
}

\lref\CachazoFWA{
  F.~Cachazo and A.~Strominger,
``Evidence for a New Soft Graviton Theorem,''
[arXiv:1404.4091 [hep-th]].
}

\lref\CasaliXPA{
  E.~Casali,
``Soft sub-leading divergences in Yang-Mills amplitudes,''
JHEP {\bf 1408}, 077 (2014).
[arXiv:1404.5551 [hep-th]].
}
\lref\BernBX{
  Z.~Bern, A.~De Freitas and H.L.~Wong,
``On the coupling of gravitons to matter,''
Phys.\ Rev.\ Lett.\  {\bf 84}, 3531 (2000).
[hep-th/9912033].
}
\lref\WeinbergKQ{
  S.~Weinberg and E.~Witten,
  ``Limits on Massless Particles,''
Phys.\ Lett.\ B {\bf 96}, 59 (1980).
}

\lref\StiebergerHZA{
  S.~Stieberger and T.R.~Taylor,
``Superstring Amplitudes as a Mellin Transform of Supergravity,''
Nucl.\ Phys.\ B {\bf 873}, 65 (2013).
[arXiv:1303.1532 [hep-th]];
``Superstring/Supergravity Mellin Correspondence in Grassmannian Formulation,''
Phys.\ Lett.\ B {\bf 725}, 180 (2013).
[arXiv:1306.1844 [hep-th]].
}

\lref\DvaliILA{
  G.~Dvali, C.~Gomez, R.S.~Isermann, D.~L\"ust and S.~Stieberger,
``Black Hole Formation and Classicalization in Ultra-Planckian $2 \ra N$ Scattering,''
[arXiv:1409.7405 [hep-th]].
}

\Title{\vbox{\rightline{MPP--2015--001}
}}
{\vbox{\centerline{Graviton Amplitudes from Collinear Limits}
\medskip\centerline{of Gauge Amplitudes}
}}
\medskip
\centerline{Stephan Stieberger$^a$ and Tomasz R. Taylor$^b$}
\bigskip
\centerline{\it $^a$ Max--Planck--Institut f\"ur Physik}
\centerline{\it Werner--Heisenberg--Institut, 80805 M\"unchen, Germany}
\medskip
\centerline{\it  $^b$ Department of Physics}
\centerline{\it  Northeastern University, Boston, MA 02115, USA}

\vskip15pt

\medskip
\bigskip\bigskip\bigskip
\centerline{\bf Abstract}
\vskip .2in
\noindent

\noindent
We express all tree-level graviton amplitudes in Einstein's gravity  as the collinear limits of a linear combination of pure Yang-Mills amplitudes in which each graviton is represented by two gauge bosons, each of them carrying exactly one half of graviton's momentum and helicity.

\Date{}
\noindent
\goodbreak
\break

\noindent
The amplitudes describing scattering processes of many gravitons in quantized Einstein's \break general relativity are related to the amplitudes describing vector gauge boson scattering in Yang-Mills theory. As discovered by Kawai, Lewellen and Tye (KLT) \KawaiXQ\ in 1985,  gravitational amplitudes can be written as ``squares'' of gauge amplitudes, more precisely as bilinear forms of partial gauge amplitudes weighted by the kinematic coefficients presently known as the ``KLT kernel.'' Kawai, Lewellen and Tye discovered these relations in the framework of string theory, as a connection between closed and open string amplitudes. Their quadratic form is a direct consequence of the factorization of the graviton vertex operator into the operators creating non-interacting left- and right-moving fluctuations of the world-sheet. They support an intuitive picture of the closed string as a loop of two open strings connected at the ends \SiegelSK. More recently, Cachazo, He and Yuan \refs{\CachazoHCA,\CachazoIEA,\CachazoNSA}, developed a novel representation of graviton amplitudes, by utilizing the so-called scattering equations. Here again, gravity appears as a ``square'' of gauge theory.

In a recent work \StiebergerCEA, we expressed tree-level Einstein-Yang-Mills amplitudes involving one graviton and an arbitrary number of gauge bosons as {\it linear\/} combinations of pure Yang-Mills amplitudes in which the graviton appears as a pair of collinear vector bosons, each of them carrying exactly one-half of its momentum and helicity. This result is a low energy, field theory manifestation of a much broader class of linear relations between closed and open string amplitudes, which will be discussed elsewhere \stnew.
It indicates that, in some way, the graviton can be considered as a pair of gauge bosons beyond the world--sheet, as a bound state in physical space-time. Although Weinberg-Witten theorem \WeinbergKQ\ rules out a massless spin 2 graviton emergent from pure gauge dynamics\foot{This is due to the existence of a Lorentz-covariant energy-momentum tensor in Yang-Mills theory.}, it is possible that such linear relations reflect something more subtle.

In the present work, we express the $N$--graviton amplitude  as a collinear limit of a particular linear combination of pure Yang-Mills (partial) 2$N$--gluon\foot{For brevity vector gauge  bosons are called ``gluons'' below. All Yang-Mills amplitudes are partial, associated to a single trace of (arbitrary) gauge group generators in the fundamental representation.}  amplitudes. The paper is organized in a simple way. First, we establish notation. Then we will state the result and prove it by showing agreement with the KLT formula. As an illustration, we will work out explicitly the three--graviton case.

The momenta $k_i$ and helicities $\lambda_i=\pm 2$ of $N{-}1$ gravitons will be split into momenta $p_i, q_i$ and helicities $\mu_i, \nu_i$ of $2(N{-}1)$ gluons in the following way:
\eqn\split{
p_i=q_i={k_i\over 2}~,\qquad \mu_i=\nu_i={\lambda_i\over 2}~,\qquad i=1,2,\dots,{N{-}1}.}
We will also introduce one additional pair of gluons, with the momenta $p$ and $q$, and {\it opposite\/} helicities, $\mu={+}1$ and $\nu={-}1$, respectively. With all momenta assumed to be incoming into the scattering process, the momentum conservation is
\eqn\cons{
\sum_{i=1}^{N{-}1}p_i+\sum_{i=1}^{N{-}1}q_i+p+q=\sum_{i=1}^{N{-}1}k_i+p+q=0.}
All momenta are on-shell (light-like).
It is convenient to represent $p$ and $q$ as matrices which factorize into helicity spinor variables in the following way:
\eqn\split{/\hskip -2mm p~=\sigma_p\tilde\sigma_p~,\qquad/\hskip -1.7mm q~=\sigma_q\tilde\sigma_q~.}

{}For real light-like momenta $p$ and $q$, the limit $s_{pq}= (p+ q)^2=\langle pq\rangle[qp]\to 0$ at finite $p$ and $q$ constraints the respective three-momenta to a collinear configuration, {\it i.e}.\ both pointing in the same direction\foot{We are using standard notation of the helicity formalism \notation.}. Here, however, we will be considering complex momenta, which allows two ways of reaching $s_{pq}= 0$:
\eqn\limh{[pq]\to 0~{\rm with}~ \langle pq\rangle\neq 0~:\qquad\tilde\sigma_p\to x\tilde\sigma_q~,}
where  $x$ is an arbitrary number, and similarly,
\eqn\limah{\langle pq\rangle\to 0~{\rm with}~ [pq]\neq 0~:\qquad\sigma_p\to x\sigma_q~.}

We will be using the following $2N$--gluon Yang-Mills partial amplitudes
\eqn\adef{\eqalign{
A[&p,N{-}1,1,\pi(2,3,\ldots,N{-}2),1,\rho(2,\ldots,N{-}2),N{-}1,q]
\cr &\equiv
A[p,\mu={+}1;p_{N{-}1},\mu_{N{-}1};\dots;p_{(N{-}2)_\pi},\mu_{(N{-}2)_\pi}; q_1,\nu_1;\dots;q_{N{-}1},\nu_{N{-}1};
q,\nu={-}1]\ ,}}
where $\pi,\rho\in S_{N{-}3}$ denote permutations of $N{-}3$ elements and  $i_\pi\equiv\pi(i)$, $j_\rho\equiv\rho(j)$.
We will also need the KLT kernel $S[\pi|\rho]$ introduced in
\refs{\KawaiXQ,\BernSV,\Bohr}
\eqn\kernel{
S[\pi|\rho]\equiv S[\, \pi(2,\ldots,N-2) \, | \, \rho(2,\ldots,N-2) \, ] = \prod_{i=2}^{N-2} \Big( \, s_{1,i_\pi} \ + \ \sum_{j=2}^{i-1} \theta(i_\pi,j_\pi) \, s_{i_\pi,j_\pi} \, \Big)\ ,}
where $s_{i,j}\equiv (p_i+p_j)^2$ and  $\theta(i_\pi,j_\pi)=1$
if the ordering of the legs $(i_\pi,j_\pi)$ and $(i_\rho,j_\rho)$ is the same for
$\pi(2,\ldots,N-2)$ and $\rho(2,\ldots,N-2)$, and zero otherwise\foot{Note, that the kernel \kernel\
does not  depend explicitly on the momenta $p$ and $q$.}.

\medskip
\noindent
{\bf Theorem}:
The $N$--graviton amplitude in Einstein's gravity  is given at the tree level by:
\eqn\aplus{\eqalign{
A_E[k_1,&\lambda_1;\dots;k_{N{-}1}, \lambda_{N{-}1};k_N=p+q,\lambda_N=+2]=
\lim_{[pq]\to 0}\bigg({1\over 2x}\bigg)^4{[pq]\over\langle pq\rangle}s_{pq}^2\cr
&\times\sum_{\pi,\rho\in S_{N-3}}S[\pi|\rho]\
A[p,N{-}1,1,\pi(2,3,\ldots,N{-}2),1,\rho(2,\ldots,N{-}2),N{-}1,q]\ ,}}
where  the limit is defined in Eq. \limh. 
Similarly,
\eqn\aminus{\eqalign{
A_E[k_1,&\lambda_1;\dots;k_{N{-}1}, \lambda_{N{-}1};k_N=p+q,\lambda_N=-2]
=\lim_{\langle pq\rangle\to 0}\big( 2x\big)^4{\langle pq\rangle\over [pq]}s_{pq}^2\cr
&\times
\sum_{\pi,\rho\in S_{N-3}}S[\pi|\rho]\
A[p,N{-}1,1,\pi(2,3,\ldots,N{-}2),1,\rho(2,\ldots,N{-}2),N{-}1,q]\ ,}}
with the limit defined\foot{In the above relations, we omit constant factors involving Yang-Mills and gravitational coupling constants.}  in Eq. \limah.

\medskip 
\noindent
{\bf Proof}: In order to prove Eq. \aplus, we note that $[pq]s_{pq}^2\sim s_{pq}^3$, therefore the limit $[pq]\to 0$ pushes the Yang Mills amplitude on the r.h.s.\ of \aplus\ into a highly singular (triple ``factorization pole'') kinematic configuration. In the first step, we factorize on the pole in the $N$-gluon channel shown in Fig.1, with the total momentum of:
\eqn\fact{
l=\sum_{i=1}^{N{-}1}p_i+p={p-q\over 2}~,\qquad l^2=-{s_{pq}\over 4}\ .}
\centerline{\epsfxsize=0.5\hsize\epsfbox{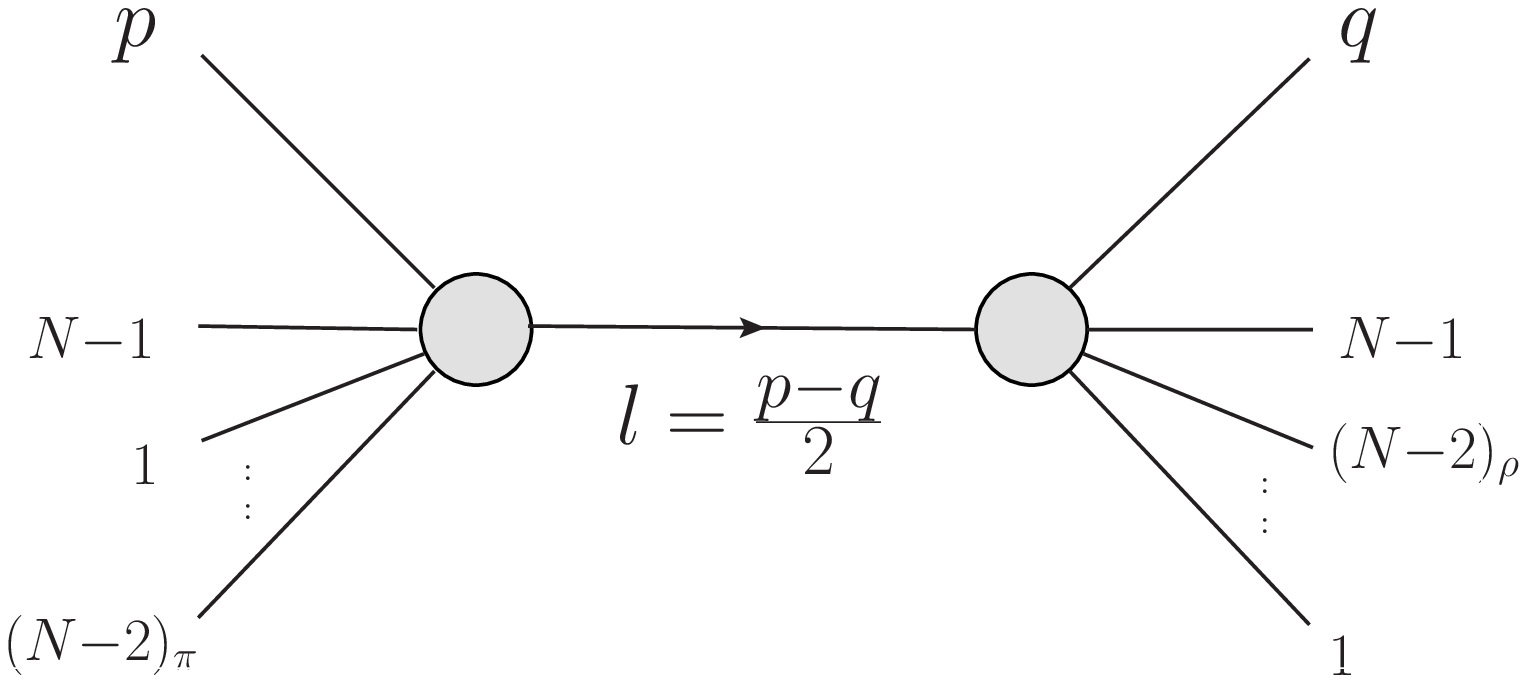}}
\noindent{\ninepoint\sl \baselineskip=8pt \centerline{{\bf Figure 1}: \sl
Factorization in the $N$-gluon channel.}}\medskip\noindent
Furthermore, the subamplitude on the left side of Fig.1 develops a pole in the two-gluon channel with
\eqn\pn{p_N=p-l={p+q\over 2}={k_N\over 2}~,\qquad p_N^2={s_{pq}\over 4}~,}
as shown in Fig.2.  Similarly, the subamplitude on the right side of Fig.1 develops a pole in the two-gluon channel with
\eqn\qn{q_N=q+l={p+q\over 2}={k_N\over 2}~,\qquad q_N^2={s_{pq}\over 4}~.}
\centerline{\epsfxsize=0.5\hsize\epsfbox{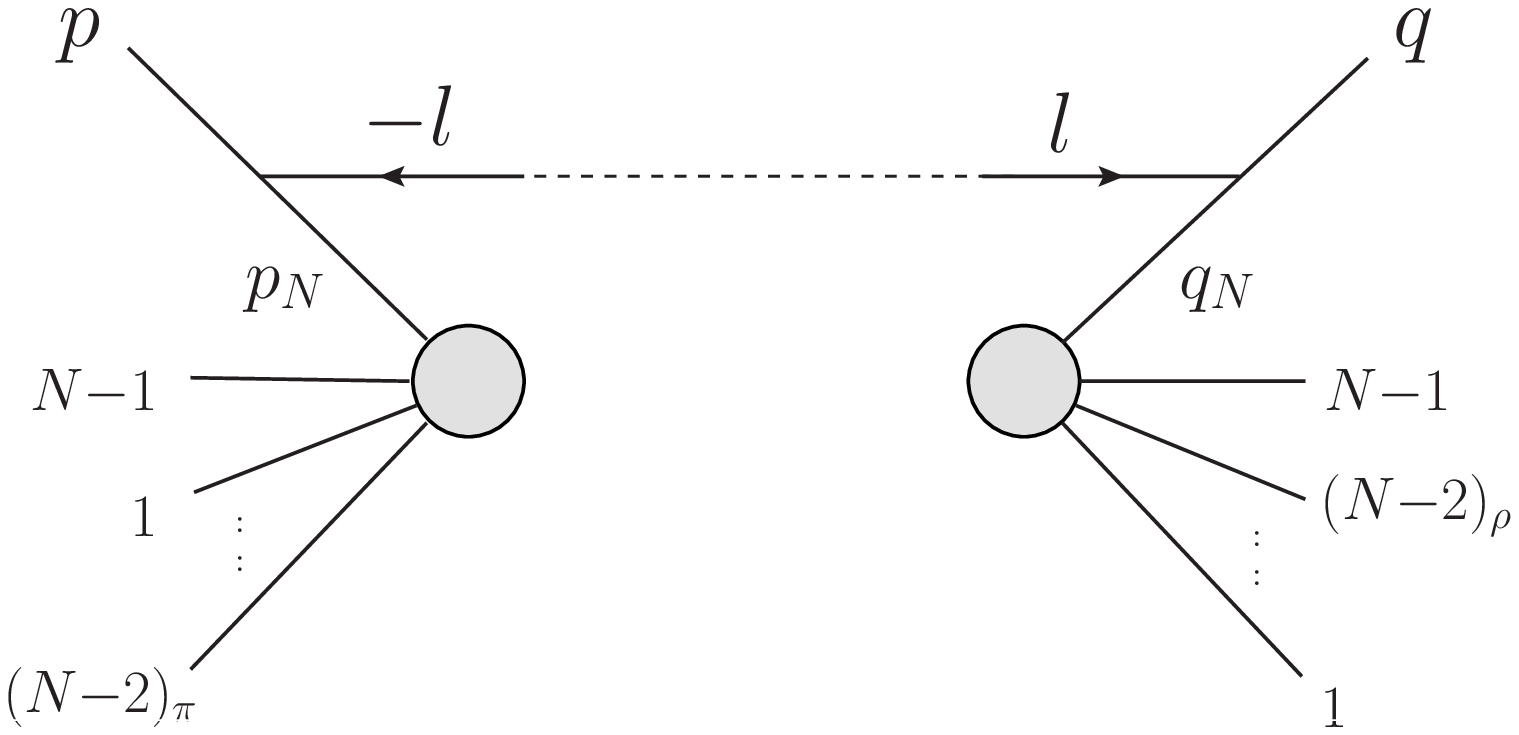}}
\noindent{\ninepoint\sl \baselineskip=8pt \centerline{{\bf Figure 2}: \sl
Factorization of subamplitudes.}}
It is easy to show that in the limit \limh, there is a unique helicity configuration contributing to the triple pole
\eqn\triple{\eqalign{
A[p,N{-}1,1,&\pi(2,3,\ldots,N{-}2),1,\rho(2,\ldots,N{-}2),N{-}1,q]~\rightarrow~ \bigg( {4\over s_{pq}}\bigg)^3\times\cr
\times &A[p^+,-l^-,-p_N^-]\times A[p_N,\mu_N={+}1;N{-}1,1,\pi(2,3,\ldots,N{-}2)]\cr
\times &A[1,\rho(2,\ldots,N{-}2),N{-}1;q_N,\nu_N={+}1]\times A[q^-,l^+,-q_N^-]~,
}}\smallskip\noindent
where we used $\pm$ superscripts to indicate the respective $\pm 1$ vector boson helicities. In this limit, the three-gluon amplitudes reduce to:
\eqn\three{A[p^+,-l^-,-p_N^-]={x^3\over 2}\langle pq\rangle~,\qquad
A[q^-,l^+,-q_N^-]={x\over 2}\langle pq\rangle~.}
After inserting Eqs. \triple\ and \three\ into Eq. \aplus, we obtain
\eqn\ares{\eqalign{
A_E[k_1,\lambda_1;\dots;k_{N{-}1},\lambda_{N{-}1};&k_N,\lambda_N=+2]\cr
=\sum_{\pi,\rho\in S_{N-3}}&S[\pi|\rho]\ A[p_N,\mu_N={+}1;N{-}1,1,\pi(2,3,\ldots,N{-}2)]\cr
&\times A[1,\rho(2,\ldots,N{-}2),N{-}1;q_N,\nu_N={+}1]\ ,}}
which is the KLT formula for the $N$-graviton amplitude \refs{\KawaiXQ,\BernSV}. The proof of
Eq. \aminus\ proceeds in a similar way\foot{The amplitudes involving string dilatons can be obtained in a similar way, as collinear limits of Yang-Mills amplitudes involving two additional gauge bosons, but carrying identical helicities, {\it i.e}.\ $\mu=\nu=\pm 1$.}.

\medskip
\noindent
{\bf Example}: Three--graviton amplitude.

\noindent
{}For $N=3$, the KLT kernel is trivial and Eq. \aplus\ reads
\eqn\exam{
A_E[k_1,\lambda_1;k_2,\lambda_2;k_3=p+q,\lambda_3=+2]
=\lim_{[pq]\to 0}\bigg({1\over 2x}\bigg)^4{[pq]\over\langle pq\rangle}s_{pq}^2
A[p,2,\{1,1\},2,q]~,}
where we used curly brackets to indicate symmetrization in $\{p_1,q_1\}$ before setting $p_1=q_1$. The symmetrization removes the collinear singularity at $p_1=q_1$; it is necessary in the $N=3$ case only. The r.h.s.\ of Eq.\exam\ can be rewritten by using the BCJ relation \BernQJ
\eqn\exambcj{
s_{pq}^2\ A[p,2,\{1,1\},2,q]=s_{pq}\ \big( -s_{pq}\ A[p,1,2,1,2,q]+2\; s_{q1}\ A[p,2,1,2,q,1]\ \big)~,}
which is manifestly finite.
The combination on the r.h.s.\ of the above equation appears in the zero string slope limit of the four--particle open--closed string disk amplitude involving two gravitons and two gauge bosons, {\it cf}.\ Eq. (3.40) of Ref. \StiebergerHQ:
\eqn\examb{
s_{pq}^2\ A[p,2,\{1,1\},2,q]= A[k_1,\lambda_1;k_2,\lambda_2;p,\mu;q,\nu]~.}
In this way, we find that the limit of Eq. \exam\ amounts to factorizing this Einstein-Yang-Mills amplitude in the $s$ channel, on the graviton pole. In Ref. \stnew, we will show that the same conclusion holds for higher $N$: the triple pole limits of Yang-Mills amplitudes in Eqs. \aplus\ and \aminus\ correspond to the degeneration limit of disk amplitudes involving $N{-}1$ gravitons (closed strings) and two gauge bosons, {\it i.e.} open strings attached to the disk boundary. In this limit, the boundary shrinks to a point and the amplitude factorizes into $N$-graviton amplitude on the sphere times the amplitude for one of the gravitons to decay into a pair of gauge bosons. Now returning to the case of $N=3$, we see from Eq. \exambcj\ that a non-vanishing amplitude requires two gravitons to carry opposite helicities, thus we set
$\lambda_1={+}2,~\lambda_2={-}2$. The amplitude \examb\ can be computed by substituting the well-known six-point NMHV Yang-Mills amplitudes \notation\ into the r.h.s.\ of Eq. \exambcj, giving
\eqn\examp{A[k_1,\lambda_1={+}2;k_2,\lambda_2={-}2;p,\mu={+}1;q,\nu={-1}]={[1p][1q]\over
\langle 1p\rangle\langle 1q\rangle}{\langle 2q\rangle^4\over s_{pq}}
~.}
Next, we substitute it to Eqs.\examb\ and \exam, and take the limit by using
\eqn\sxub{x={[1p]\over [1q]}={[2p]\over [2q]}~,\qquad k_3=p+q=-k_1-k_2~,}
which yields the correct result:
\eqn\xfin{A_E[k_1,\lambda_1={+}2;k_2,\lambda_2={-}2; k_3,\lambda_3={+}2]={[13]^6\over
[12]^2[23]^2}~.}

To summarize, we proposed a new representation of gravitational amplitudes at the tree level. It would be interesting to learn whether it can be extended beyond the tree level, to improve our understanding of loop corrections in quantum gravity.

\vskip0.5cm
\goodbreak
\leftline{\noindent{\bf Acknowledgments}}

\noindent
This material is based in part upon work supported by the National Science Foundation under Grant No.\ PHY-1314774.   Any
opinions, findings, and conclusions or recommendations expressed in
this material are those of the authors and do not necessarily reflect
the views of the National Science Foundation.

\listrefs

\end
{\it Since $p_1=q_1$ (with $\mu_1=\nu_1$) for adjacent gluons, in order to obtain a well-defined amplitude, we start from $p_1\neq q_1$ and take the limit after symmetrizing in $\{p_1,q_1\}$, thus removing the leading collinear singularity. By using BCJ relations \BernQJ, such a symmetric combination can always be rewritten in terms of manifestly finite amplitudes in which the collinear gluons appear at non--adjacent positions, {\it cf}.\ in the example discussed at the end of the paper.} {\bf Is this paragraph still an issue - only for $N=3$, so perhaps we can move this comment to the example ?}
\eqn\adef{\eqalign{A[&p,N{-}1,N{-}2,\dots,2, 1,1,2,\dots,N{-}2,N{-}1,q]\cr &\equiv
A[p,\mu={+}1;p_{N{-}1},\mu_{N{-}1};\dots;p_1,\mu_1; q_1,\nu_1;\dots;q_{N{-}1},\nu_{N{-}1};
q,\nu={-}1]~.
}}